\documentclass[10pt,preprint]{aastex}
\usepackage{emulateapj5}

\newcommand{\lya}{Ly$\alpha$}
\newcommand{\ha}{H$\alpha$}
\newcommand{\hb}{H$\beta$}

\newcommand{\ci}{{\rm C}~{\sc i}}
\newcommand{\cii}{{\rm C}~{\sc ii}}
\newcommand{\civ}{{\rm C}~{\sc iv}}
\newcommand{\siii}{{\rm Si}~{\sc ii}}
\newcommand{\siiii}{{\rm Si}~{\sc iii}}
\newcommand{\siiv}{{\rm Si}~{\sc iv}}
\newcommand{\nii}{{\rm N}~{\sc ii}}
\newcommand{\nv}{{\rm N}~{\sc v}}
\newcommand{\oii}{{\rm O}~{\sc ii}}
\newcommand{\oiii}{{\rm O}~{\sc iii}}
\newcommand{\mgii}{{\rm Mg}~{\sc ii}}
\newcommand{\alii}{{\rm Al}~{\sc ii}}
\newcommand{\feii}{{\rm Fe}~{\sc ii}}
\newcommand{\zem}{$z_{em}$}
\newcommand{\zabs}{$z_{abs}$}
\newcommand{\kms}{km~s$^{-1}$}
\newcommand{\kp}{$K^{\prime}$}
\newcommand{\nb}{$NB$}
\newcommand{\flux}{ergs~s$^{-1}$~cm$^{-2}$}
\newcommand{\p}{$^{\prime}$}
\newcommand{\pp}{$^{\prime\prime}$}
\newcommand{\fha}{$f_{H\alpha}$}
\newcommand{\spy}{M$_{\odot}$~yr$^{-1}$}

\slugcomment{To appear in the Astronomical Journal}
\shorttitle{NIR search for C IV absorption counterparts}
\shortauthors{Misawa et al.}

\begin{document}

\title{Near-Infrared search for C IV absorption counterparts along the
  line-of-sights to pair quasars$^{1,2}$}

\footnotetext[1]{Based on data collected at Subaru Telescope, which is
  operated by the National Astronomical Observatory of Japan}
\footnotetext[2]{Based on data collected at The United Kingdom
  Infrared Telescope, which is operated by the Joint Astronomy Centre
  on behalf of the U.K. Particle Physics and Astronomy Research
  Council.}

\author{
        Toru Misawa\altaffilmark{3}, 
        Nobunari Kashikawa\altaffilmark{4,5},
	Youichi Ohyama\altaffilmark{6},
        Tetsuya Hashimoto\altaffilmark{7} and
        Masanori Iye\altaffilmark{4,5,7}
        }

\altaffiltext{3}{Department of Astronomy and Astrophysics,
  Pennsylvania State University, 525 Davey Lab, University Park, PA
  16802; misawa@astro.psu.edu}
\altaffiltext{4}{National Astronomical Observatory, 2-21-1 Osawa,
  Mitaka, Tokyo 181-8588, Japan}
\altaffiltext{5}{Department of Astronomical Science, The Graduate
  University for Advanced Studies, 2-21-1 Osawa, Mitaka, Tokyo
  181-8588, Japan}
\altaffiltext{6}{Department of Infrared Astrophysics, Institute of
  Space and Astronautical Science, Japan Aerospace Exploration Agency,
  3-1-1 Yoshinodai, Sagamihara, Kanagawa, 229-8510, Japan}
\altaffiltext{7}{Department of Astronomy, School of Science,
  University of Tokyo, 7-3-1 Hongo, Bunkyo-ku, Tokyo 113-0033 Japan}

\begin{abstract}
\renewcommand{\thefootnote}{\fnsymbol{footnote}}
We carried out a Subaru and UKIRT near infrared imaging survey for
\ha\ emitting galaxies around two pair quasar systems
(Q0301-005/Q0302-003 and Q2343+125/Q2344+125), and a triple quasar
system (KP76/KP77/KP78). Narrow band near infrared filters covering
the \ha\ emission expected for galaxies at the confirmed \civ\
absorption redshift toward the quasar systems were used for this
survey. These quasar pairs or triplet are separated at most by 17
arcmins ($\sim$ 5$h^{-1}$ Mpc in proper distance) from each other on
the sky, and have common \civ\ absorption lines at almost identical
redshifts at $z$ = 2.24 -- 2.43, which suggests there could be a
Mpc-scale absorbing systems such as a cluster, or a group, of galaxies
that cover all the line-of-sights to the pair/triple quasars. Using
narrow-band deep images, we detected five candidates for H$\alpha$
emitting galaxies around two of the six fields, Q2343+125 and
Q2344+125, whose apparent star formation rates are, extremely high, 20
-- 466 \spy. However, all or most of them are not likely to be
galaxies at the absorption redshift but galaxies at lower redshift,
because of their extreme brightness. In the fields of the other
quasars, we detected no star-forming galaxies, nor did we find any
number excess of galaxy counts around them. This no-detection results
could be because the luminosities and star formation rates of galaxies
are lower than the detection limits of our observations (\kp\ $>$ 21
and $SFR$ $<$ 1.8--240$h^{-2}$\spy). They could be located outside of
the observed field around Q0301 and Q0302, since our targetting
field covers only 2\%\ of this pair quasar field. But this is not the
case for the other pair/triple quasar fields, because we got
effectively large field coverage fractions ($\sim$ 33 -- 75\%). 
Otherwise, most \civ\ absorption lines could be ascribed not to
cluster of galaxies, but to isolated star forming pockets far from
bright galaxies and could be analogous objects to weak \mgii\
absorbers.
\end{abstract}


\section{Introduction}
Several quasars, which are separated from each other on the sky by a
few arcmins, have sometimes common metal absorption lines at almost
identical redshifts (e.g., Shaver, Boksenberg, \& Robertson 1982;
Jakobsen et al. 1986; Crotts \& Fang 1998). The presence of such
common metal absorption lines implies that Mpc-scale gas clouds exist
at the redshift, and that they are covering both lines of sight to the
quasars. Since it is difficult to assume that a single homogeneous Mpc
scale absorber covers both the lines of sight to those quasar pairs
based on the framework of the prevalent CDM dominant universe, these
metal lines could be produced by gas clouds that are clustering and
forming Mpc-scale system (e.g., clouds in galaxies that are members of
a Mpc-scale cluster (group) of galaxies). Francis \& Hewett (1993)
estimated the probability of having strong \lya\ absorption lines
(i.e., analogue of metal lines) at almost same redshift in two lines
of sight separated by a few arcmins is an order of $10^{-4}$. If there
is a cluster (group) of galaxies, the probability would be
increased. Although several high-$z$ cluster of galaxies have recently
discovered, it is still observationally difficult to detect emission
lines of star forming galaxies in the redshift desert at $z$ $>$
1.5. However, observations around pair quasars with common metal
absorption systems are quite promising (e.g., Francis et
al. 1996). High-$z$ cluster of galaxies are excellent targets to
investigate star formation histories.

To date, several cluster (group) of galaxies have been detected at $z$
$>$ 2. They are often detected around radio-loud quasars or radio
galaxies. Pentricci et al. (2000) found 14 Ly$\alpha$ emitting
galaxies within the projected distance of 1.5 Mpc from the powerful
radio galaxy Q1138-262 at $z$=2.16. Pascarelle et al. (1996) also
detected two Ly$\alpha$ emitting galaxies at $z$ $\sim$ 2.39 in the
field around the weak radio galaxy, 53W002, and also confirmed them
spectroscopically. There are several other candidates for cluster
(group) of galaxies at $z$ $>$ 2 discovered by narrow-band (\nb)
imaging observations (e.g., LeF$\grave{e}$vre et al. 1996, Hu \&
McMahon 1996). Not only \lya\ but H$\alpha$ and [\oiii] are also useful
lines for identification of star forming galaxies (e.g., Teplitz,
Malkan, \& McLean 1998; Iwamuro et al., 2000). Galaxies have also been
discovered in the fields around quasars as counterparts of strong
absorption systems. \civ\ absorption systems with $W_{rest}$
(rest-frame equivalent width) $>$ 0.4 \AA\ and \mgii\ absorption
systems with $W_{rest}$ $>$ 0.3 \AA\ are thought to have $\sim$70 and
$\sim$40 kpc sizes around $L^*$ galaxies, by comparing Press-Schechter
function (as luminosity function of galaxy) and the number densities
of these absorption systems per redshift. Charlton \& Churchill (1996)
showed that both spherical halo and randomly oriented disks with only
70--80\% covering factors of gas clouds can recover the observed
properties of \mgii\ absorbers, by performing a Monte Carlo
simulation. This means that the distribution of the \mgii\ absorbers
around galaxies are not smooth but patchy. There are some galaxy
surveys around {\it single} quasars (e.g., Bergeron \& Boisse 1991;
Steidel, Dickinson, \& Persson 1994; Lanzetta et al. 1998; Chen \&
Lanzetta 2001; Chen et al. 2001). However, for {\it pair} quasar
regions, only a few observations have been carried out based on common
metal absorption lines of pair quasars (e.g., Teplitz et al., 1998;
Francis, Woodgate, \& Danks 1997; Francis et al., 1996), in spite of
its high potential.

In this paper, we report the results of our near-infrared (NIR) \nb\
imaging survey of the fields around pair/triple quasars that
have common metal absorption lines at $z$ $\sim$ 2.3. They are
separated at most by 17 arcmins ($\sim$ 5$h^{-1}$ Mpc in proper
distance, with $h$=$H_0$/72 km~s$^{-1}$~Mpc$^{-1}$). Our objectives
are to search for star forming galaxies that produce the common metal
absorption lines, and see if there are galaxies that are member of the
cluster (or group) of galaxies including those star forming galaxies.

We present the outline of the observations and data reduction in $\S$
2, and the brief description of the photometric analysis in $\S$ 3. In
$\S$ 4, we present the result for each quasar field. We summarize and
discuss our results in $\S$ 5. Throughout this paper, we assume
$H_{0}$=72 km~s$^{-1}$~Mpc$^{-1}$, $\Omega_0$ = 0.3,
$\Omega_{\Lambda}$ = 0.7, and $q_{0}$=0.5.

\section{Observation and Data Reduction}
We observed the fields of pair/triple quasars. These quasars have
common absorption systems (at least contain \civ\ doublets in them)
with small redshift difference, $\Delta z$ $\sim$ 0.005, which
corresponds to velocity difference, $\Delta v$ $\sim$ 500 \kms, in the
frame of the absorbers. However, we should notice that if redshift
difference is caused by the Hubble flow, these absorbers would be
separated by much larger than the typical size of cluster of galaxies
along the line of sight. We chose \nb\ filters which cover redshifted
\ha\ emission lines. Filter name, central wavelength, band
width, corresponding redshift for \ha\ emission line detection, and
the bandpass ratio of \nb\ to broad-band (\kp-band), are listed in
Table 1. To see the color excess, we also carried out \kp-band imaging
observations. The observations were performed with either the Cooled
Infrared Spectrograph and Camera for OHS (CISCO; Motohara et al. 1998)
on the Subaru Telescope (Iye et al. 2004), or UKIRT First-Track Imager
(UFTI; Roche et al. 2002) on the UKIRT. Both instruments have HAWAII
1024 $\times$ 1024 pixel HgCdTe arrays that cover a field of view
(FoV) of 108\pp~$\times$~108\pp\ and 92\pp~$\times$~92\pp,
respectively. We summarize the observation logs in Table 2; columns
(1) and (2) are quasar name and its emission redshift. Absorption
redshift of common metal lines is given in column (3). Identified ion
transition is listed in column (4). The data was taken on the date in
column (6) using the filter in column (5). Exposure time and average
seeing size are in columns (7) and (8). Columns (9) and (10) are
detection limit with 3$\sigma$ and 5$\sigma$ levels, which are
magnitudes of the faintest artificial objects that are placed in the
observed frame and extracted with 3$\sigma$ or 5$\sigma$ detection
level. We will describe the simulation in detail in \S\ 3. In columns
(11) and (12) we also present 3$\sigma$ detection limits of \ha\
emission line and correspondent star formation rate. References
of spectroscopic observations are given in column (13).

Data reduction was processed in a standard manner with IRAF. All
objects are identified by the Source Extractor program (Bertin \&
Arnouts 1996) with detecting-threshold of 2.0$\sigma$. We evaluated
the magnitudes of objects in circular apertures twice as large as
seeing size.

\section{Photometric Analysis}
In the color-magnitude (CM) analysis to isolate \ha\ emitting objects,
we need an accurate evaluation of the photometric errors. Therefore we
have created 10,000 artificial stars with the seeing size of each
frame. Their magnitude distribution is homogeneous between \kp\ =
15--22 mag. We placed them in both \kp\ and \nb\ frames randomly. We
plot CM diagrams to compare the detected objects against the simulated
artificial objects. The candidates for intervening galaxies at $z$
$\sim$ 2.3 have \kp\ $-$ \nb\ color excess because their \ha\ emission
lines are redshifted into the \nb\ filter bandpass, which makes them
deviated from the distributions of the artificial objects. We regard
all the objects as candidates of \ha\ emitters, if they are deviated
toward the positive direction in the vertical axis (i.e., \kp\ $-$ \nb)
more than 3$\sigma$ from the distribution of the artificial objects in
CM diagrams. It is unlikely that the color excess is caused by other
lines whose rest-frame wavelength are shorter than \ha, such as \lya,
[\oii], or [\oiii] lines, because galaxies should be at $z$ $\sim$
17.5, 5.0, and 3.5 if these lines are redshifted into the bandpass of
\nb\ filters. Flux of these objects would be too weak to detect in our
observations. On the other hand, if candidate objects are much
brighter than the typical magnitudes of \ha\ emitting galaxies that
produce metal absorption lines; $J$$\sim$22.5, $H$$\sim$21.5, and
$K$$\sim$20.8 (Teplitz et al. 1998), these color excess could be due
to near-infrared emission lines from galaxies at lower redshift, such
as [\feii]1.257$\mu$, [\feii]1.644$\mu$, Pa$\alpha$, Pa$\beta$, and
Br$\gamma$, which is described in \S\ 5.

If we assume that the color excess is attributed to \ha\ emission
lines, we can estimate the flux of \ha\ line from the \kp\ and \nb\
magnitudes by
\begin{equation}
K' - NB = 2.5 \log \left(1 \pm 10^{\frac{K' - \gamma}{2.5}} \right) +
          K'_{0} - NB_{0},
\end{equation}
where $K^{\prime}_{0}$ and $NB_{0}$ are magnitude zeropoints (i.e., a
magnitude corresponding to the flux, one count per second on each
pixel) for \kp\ and \nb\ filters (e.g., Iwamuro et al. 2000). Here,
the constant, $\gamma$, is defined as follows,
\begin{equation}
\gamma = K'_{0}-2.5 \log \left( \frac{f_{H\alpha}}{W_{NB}} \right),
\end{equation}
where $W_{NB}$ is the band width of the \nb\ filter and \fha\ is a
total \ha\ line flux (without the continuum flux) that is covered by
the \nb\ filters. We define 1, 2, and 3 $\sigma$ deviation borders in 
CM diagram as the equation (1) with fixed $\gamma$-value that cover 
68.3, 95.5, and 99.7 \% of all the artificial objects.

From the luminosity of \ha\ emission line, $L_{H\alpha}$ (ergs
s$^{-1}$), we can also evaluate star formation rate ($SFR$) using the
conversion relation described in Kennicutt (1998),
\begin{eqnarray}
SFR_{H\alpha}~(total) & = & \frac{L_{H\alpha}}{1.27\times10^{41}
                            ergs~s^{-1}}~M_{\odot}~yr^{-1},
\end{eqnarray}
where we assume the Salpeter initial mass function with lower and
upper mass cutoffs of 0.1 and 100 $M_{\odot}$.

\section{Results}
The color-magnitude analysis of all the objects detected in the six
fields around pair/triple quasars yielded two and three
candidates for star forming galaxies around Q2343+125 and Q2344+125,
respectively. In this section, we describe the result for each quasar
field.

\subsection{KP76/KP77/KP78}
The quasar triplet (KP76:Q1623+2651A at \zem\ = 2.467, KP77:Q1623+2653
at \zem\ = 2.526, and KP78:Q1623+2651B at \zem\ = 2.605) is located on
the sky within a small FoV of 3 arcmin: 147\pp\ between KP76 and KP77,
127\pp\ between KP76 and KP78, and 177\pp\ between KP77 and KP78
(e.g., Crotts \& Fang 1998). All of them have \civ\ absorption lines
at $z$ $\sim$ 2.24 in their spectra with total equivalent widths,
$W_{rest}$ $\sim$ 0.14, 0.08, and 2.34 \AA, respectively. Other
transitions of metal lines such as \ci, \cii, \siii, \siiii, \siiv,
\nv\ are also identified. The radial velocity separations of these
absorption lines are within 500 \kms\ of each other in the frame of
the absorbers. The linear angular distance on the sky between these
systems corresponds to $\sim$ 1$h^{-1}$ Mpc at $z$ $\sim$ 2.24, which
is comparable to the typical size of cluster of galaxies in the local
universe.

We carried out two deep imaging observations of the fields around KP76
with UKIRT + UFTI with FoV of 90\pp~$\times$~90\pp\ (i.e.,
500$h^{-1}$~kpc~$\times$~500$h^{-1}$~kpc), and KP77/KP78 with Subaru +
CISCO with FoV of 108\pp~$\times$~108\pp\ (i.e.,
600$h^{-1}$~kpc~$\times$~600$h^{-1}$~kpc). Although we found two
objects around KP76 that were detected only in \nb\ filters, they are
confirmed to be ghost images of brightest sources in the frames. There
are not any other candidates for star forming galaxies with $SFR$ $>$
6.8$h^{-2}$~\spy\ (\fha\ $>$ 2.4$\times$10$^{-17}$ \flux) in KP76
field or $SFR$ $>$ 1.8$h^{-2}$~\spy\ (\fha\ $>$ 6.3$\times$10$^{-18}$
\flux) in KP77/KP78 field, with 3$\sigma$ detection limit of \kp\
$\leq$ 21.

\subsection{Q0301-005/Q0302-003}
This pair quasar is separated by about 17\p\ from each other on the
sky, which corresponds to $\sim$ 5.5$h^{-1}$ Mpc at $z$ $\sim$ 2.43
(e.g., Dobrzycki \& Bechtold 1991). They have common \civ\ and/or
\siiv\ doublets at $z$ $\sim$ 2.96, 2.72, and 2.43 (Cowie et al.,
1995; Songaila 1998; Steidel 1990), of which only \ha\ emissions at
$z$ $\sim$ 2.43 can be identified by the \nb\ filter of our
observation. We took two images with UKIRT + UFTI around both quasars
by putting them at the centers of frames. We identified no candidates
for star forming galaxies with $SFR$ $>$ 236$h^{-2}$~\spy\ (\fha\ $>$
6.9$\times$10$^{-16}$ \flux) in Q0301 field and $SFR$ $>$
23$h^{-2}$~\spy\ (\fha\ $>$ 6.7$\times$10$^{-17}$ \flux) in Q0302
field. The 3$\sigma$ detection limits are \kp\ = 20.1 around Q0301-005
and \kp\ = 21.1 around Q0302-003.

\subsection{Q2343+125/Q2344+125}
Q2343+125 has \civ\ absorption lines at $z$=2.4285 and 2.4308, while
Q2344+125 has corresponding strong \civ\ absorption lines at
$z$=2.4265 and 2.4292 (Sargent, Boksenberg, \& Steidel 1988). The
velocity separation of these absorption lines along the line of sight
is smaller than 400 \kms. These quasars are separated from each other
only by 5\p\ ($\sim$ 1.6$h^{-1}$ Mpc) at $z$ $\sim$ 2.43. The \civ\
absorption system in Q2343+125, which was classified as Damped \lya\
system (Lu, Sargent, \& Barlow 1998), has other metal lines such as
\alii\ $\lambda$1670, \feii\ $\lambda$1608, \siii\ $\lambda$1526.

To date, several deep imaging observations have been carried out for
this field. Bergvall et al. (1997) took an optical narrow-band deep
images to search \lya\ emitting objects around both of the quasars,
but did not find any candidates. \lya\ emissions, however, are
strongly affected by dust extinction, which makes it difficult to
detect. Therefore, Teplitz et al. (1998) took deep NIR images with a
small FoV (38\pp $\times$ 38\pp) around Q2343+125 to search H$\alpha$
emitters, but nothing was detected. Bunker et al. (1999) also carried
out a long-slit K-band spectroscopic search for \ha\ emitters in the
vicinity (within 11\pp~$\times$~2.$\!\!$\pp5) of Q2343+125, but found
nothing above 3$\sigma$ limit (\fha\ = 6.5--16$\times$10$^{-17}$
\flux).

In our NIR images taken with Subaru + CISCO, we detected two
candidates (objects A and B) for star forming galaxies at $z$ $\sim$
2.43 around Q2343+125 (Figure 1) and three candidates (objects C, D,
and E) around Q2344+125 (Figures 2), whose \kp\ magnitude, \ha\ flux,
and $SFR$ are listed in Table 3, by assuming that they are actually
galaxies at $z$ $\sim$ 2.43. Unfortunately, it is difficult to tell
the morphological type of the identified objects because of low
spatial resolutions in the observed images. The 3$\sigma$ detection
limits of these images are \kp\ = 20.5 and 21.1, respectively. These
objects were outside of the frame (or just at the borders of FoV) in
the previous observation by Teplitz et al. (1998). The CM diagrams are
also presented in Figures 3 and 4. All candidates are bright (\kp\ $<$
18.3) and their apparent $SFR$s are very large ($>$
20$h^{-2}$~\spy). We will discuss them in the next section.

% INSERT Figures 1 & 2
\begin{figure*}[t]
\centerline{
\includegraphics[width=3.in]{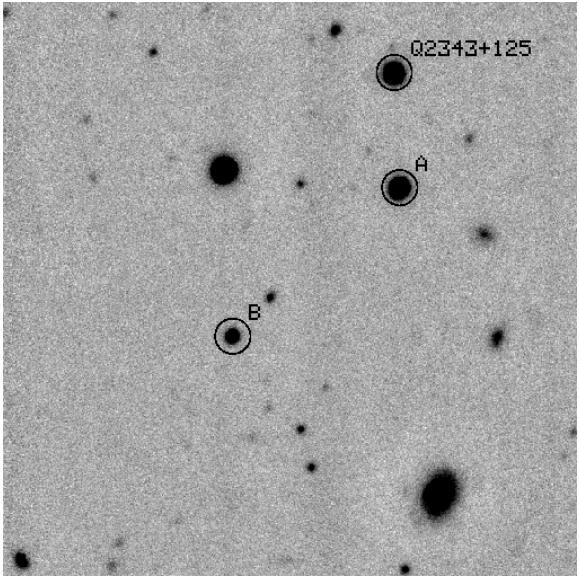}
\hspace{1cm}
\includegraphics[width=3.in]{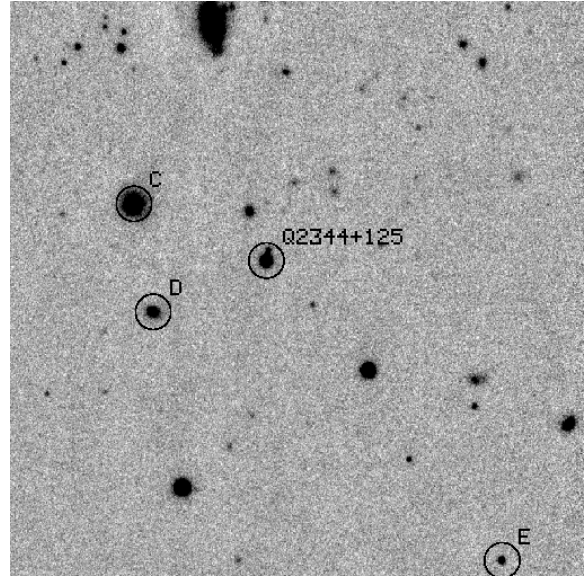}
}
\caption[Figure 1]{(Left) \kp\ image of the field around
  Q2343+125. The quasar and 2 objects that have color excess in \nb\
  filter are surrounded by circles. FoV of the images is
  $94.\!\!^{\prime\prime}5$~$\times$~$94.\!\!^{\prime\prime}5$
  (slightly trimmed from the observed image), which corresponds to
  512$h^{-1}$~kpc~$\times$~512$h^{-1}$~kpc at $z$ $\sim$ 2.43.}
\caption[Figure 2]{(Right) Same as Figure 1, but for the field around
  Q2344+125.}
\end{figure*}

% INSERT Figure 3
\begin{figure*}[b]
\includegraphics[bb=-20 17 400 290]{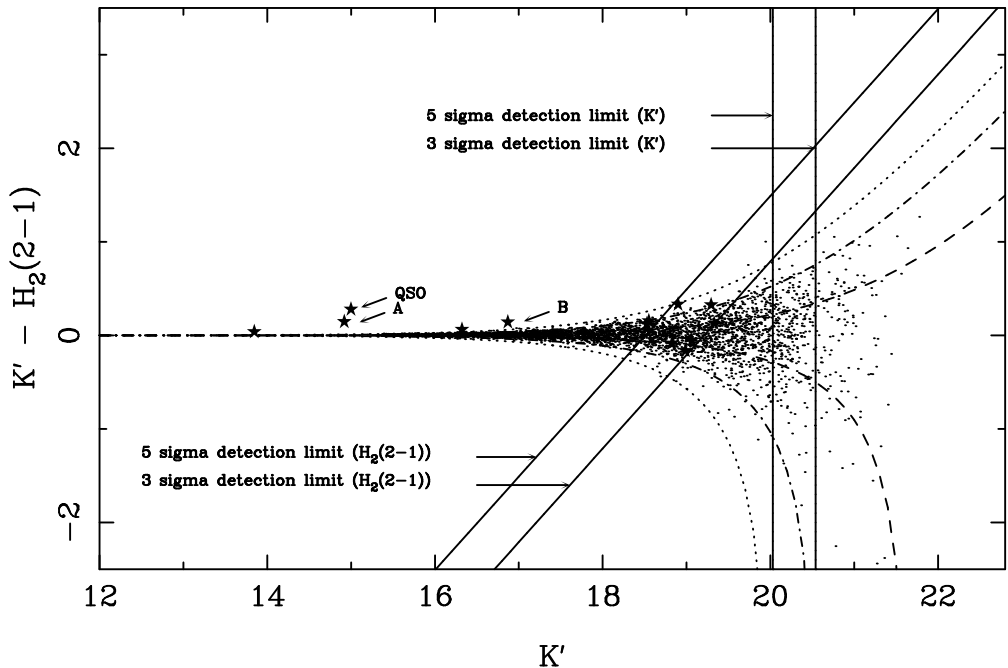}
\caption[Figure 3]{\kp $-$ \nb\ ($H_{2}$(2-1)) color versus \kp\
  diagram for the field around Q2343+125. The filled stars denote the
  objects detected in both \kp\ and \nb\ images, while the thin dots
  represent the simulated data (only one of every ten data are plotted
  to make them easy to see). The three marked objects are a quasar and
  candidates for H$\alpha$ emitting galaxies at $z$ $\sim$
  2.43. Dashed, dot-dashed, and dotted curves are 1, 2, and 3 $\sigma$
  deviations of the artificial objects placed in CM diagram. Solid
  lines denote 3$\sigma$ and 5$\sigma$ detection limits in \kp\ and
  \nb\ frames.}
\end{figure*}

% INSERT Figure 4
\begin{figure*}[b]
\includegraphics[bb=-20 17 400 290]{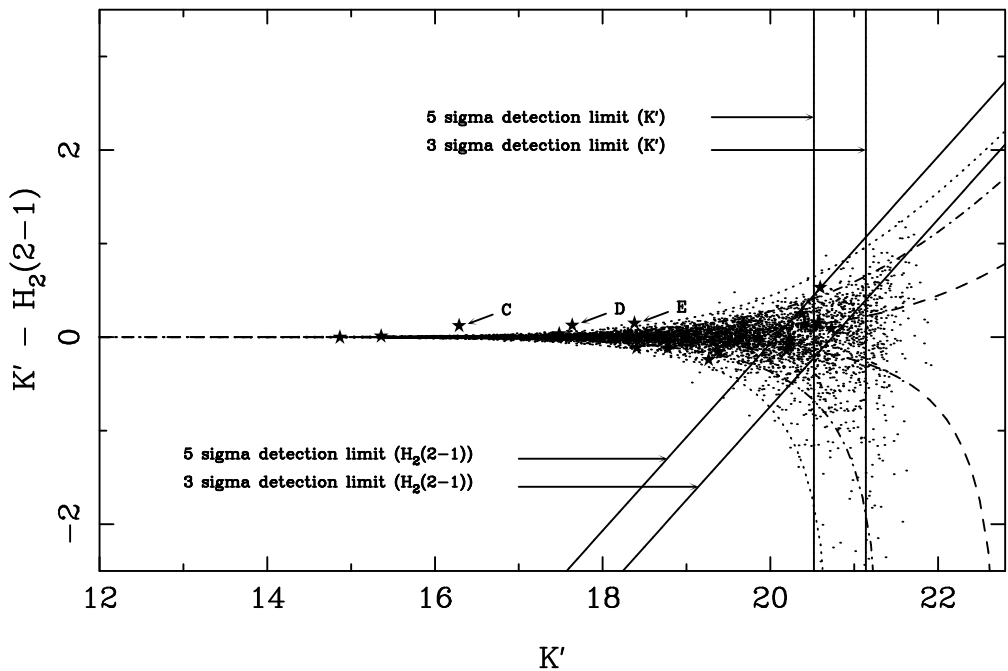}
\caption[Figure 4]{Same as Figure 3, but for the field around
  Q2344+125. Quasar is not plotted, because we do not plot objects
  blending with neighbors, and that the quasar blends.}
\end{figure*}

\section{Summary and Discussion}
We detected five candidates (objects A -- E) for intervening galaxies
around two of six pair/triple quasar fields, by NIR \nb\
imaging observations. They are all bright (\kp\ $<$ 18.3) 
with large \ha\ fluxes (\fha\ $>$ 5.7$\times$ $10^{-17}$~\flux).

Teplitz et al. (1998) detected 13 \ha\ emitters, using same method to
ours, of which the brightest two objects (\kp=15.42 with \fha\ = 249
$\times$ 10$^{-17}$ \flux\ around Q0114-089 and \kp=18.15 with \fha\ =
581 $\times$ 10$^{-17}$ \flux\ around PC2149+0221) have similar
properties to ours. Teplitz et al. (1998) regarded them as Seyfert I
galaxies, because they are extremely compact and one of them had broad
emission lines. However, all (or at least substantial fraction) of our
objects with absolute magnitude, $M_B$ $<$ $-$24$+$5$\log h$, are not
likely to be Seyfert galaxies at $z$ $\sim$ 2.43, because their volume
density\footnote[1]{We evaluated this value by assuming that all five
  objects exist within 1.6$h^{-1}$ Mpc from each other along line of
  sight, which is consistent to the separation of the lines of sight
  to the pair quasars at $z$ $\sim$ 2.43.}
($\sim$5.2$h^{3}$ Mpc$^{-3}$) is much larger than the global density
of AGNs with similar luminosities at similar redshift ($\sim$10$^{-6}$
Mpc$^{-3}$; Warren, Hewett, \& Osmer 1994). Some of them could be
foreground (active) galaxies at $z$ $\sim$ 0.04, 0.20, 0.37, 0.76, and
0.79, whose Br$\gamma$, Pa$\alpha$, [\feii]1.644$\mu$, Pa$\beta$, and
[\feii]1.257$\mu$ emission lines are redshifted into the bandpass of
\nb\ filters. These are most prominent emission lines of active
galaxies in $NIR$ window (e.g., Simpson et al. 1996; Kawara, Nishida,
\& Taniguchi 1988; Goodrich, Veilleux, \& Hill 1994). Actually, Tamura
et al. (2001) found a galaxy at $z$=0.132 whose $P\alpha$ is strong,
$f_{P\alpha}$ = 3.4$\times$10$^{-17}$~\flux\ by $NB$ imaging
observation. Thus, it seems unlikely that there exists cluster (group)
of bright galaxies at $z$ $\sim$ 2.43.

In the fields of other pair/triple quasars, we detected no galaxies at
redshifts of the common \civ\ absorption lines. There are at least
four possible reasons as follows.

First, this could be because the star formation rates are too low to
detect in the observed frames. The minimum $SFR$ and \ha\ flux we can
detect in each observed frame is described in the previous section and
summarized in Table 2. Typical $SFR$s of individual field galaxies at $z$
$\sim$ 2 have been estimated to be $\sim$ 10--35\spy\ (without dust
extinction correction) by infrared imaging surveys with narrow-band
filters and spectroscopic observations (e.g., Moorwood et al. 2000;
Iwamuro et al. 2000). Similar $SFR$ is derived for Lyman break
galaxies at $z$ $\sim$ 2--3, from spectroscopic observation based on
\hb\ emission lines ($\sim$ 20--70 \spy) with one exception ($\sim$270
\spy) (Pettini et al. 1998). Juneau et al. (2005) also estimated
$SFR$s of field galaxies at $z$ $\sim$ 2 from UV continuum
luminosities with dust extinction correction, and again got similar
values ($\sim$ 30 \spy). Teplitz et al. (1998) searched star forming
galaxies at same redshift as metal absorption systems, and found 11
\ha\ emitters at $z$ = 2.3--2.5 within 250 kpc of quasar lines of
sight. Their average $SFR$ is $\sim$50 \spy. The 3$\sigma$ detection
limits of $SFR$ in Q0301/Q0302 fields are comparable or larger than
the average $SFR$ at $z$ $\sim$ 2 in the literature. In this case,
star forming galaxies could not be detected unless they have extremely
large $SFR$s. On the other hand, image depths of the other fields
are enough to detect field star forming galaxies with typical $SFR$s,
which means that absorbers corresponding to \civ\ absorption lines
could have lower $SFR$s, compared with field galaxies at same
redshift.

Second possible reason of no-detection is that typical star forming
galaxies could be faint compared to the detection limits of our
observations. At $z$ $\sim$ 2, an average \kp\ magnitude of star
forming galaxies detected based on metal absorption lines is \kp\
$\sim$ 20.8 (Teplitz et al. 1998). On the other hand, of six fields
observed, one (or four) fields were observed to provide deep
images enough to detect such faint star forming galaxies with the
5$\sigma$ (or 3$\sigma$) detectability. Although observed images are
very deep for most of the fields, these could not enough for some
fields (i.e., Q0301 and Q2343 fields). We also confirmed that there was no
number excess of galaxy counts in the fields around the pair quasars
(Figure 5), compared with the global value of field galaxies around
Subaru Deep Field (Maihara et al. 2001). Thus, we cannot yet reject
existences of cluster (group) of galaxies in our pair/triple quasar
fields, because they could contain only faint galaxies with \kp\
$\geq$ 21.

% INSERT Figure 5
\begin{figure*}[t]
\centerline{
\includegraphics[width=5.in]{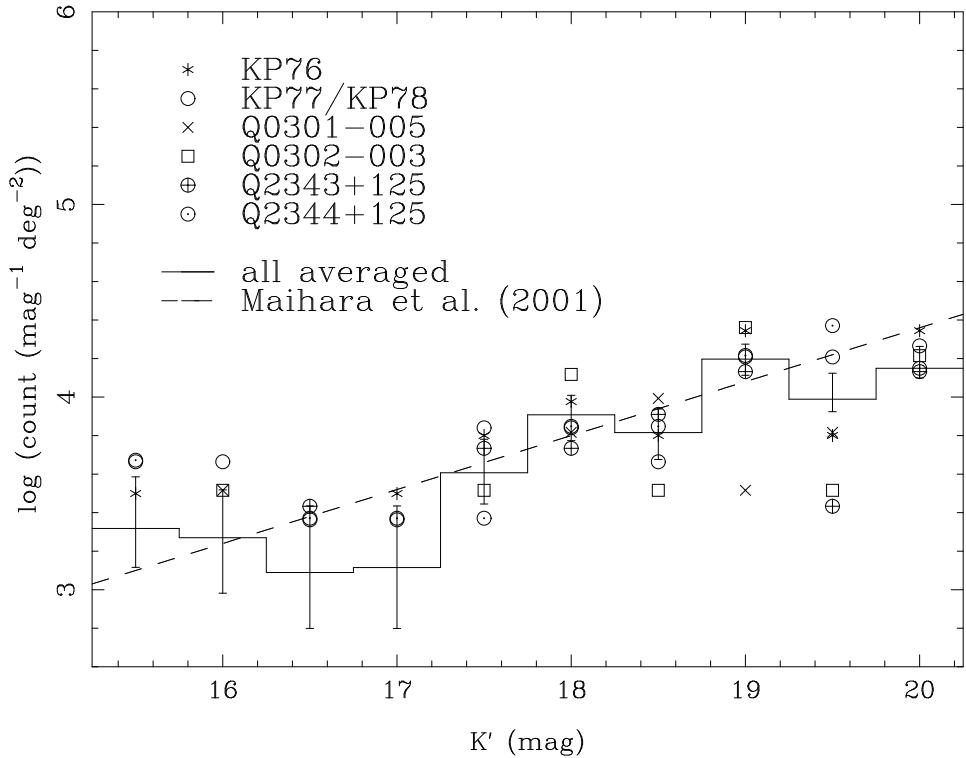}
}
\caption[Figure 5]{Galaxy number count (mag$^{-1}$ deg$^{-2}$)
  vs. \kp\ magnitude in each pair quasar field. We do not plot marks
  if no objects were found in a given bin of \kp\ magnitude in some
  quasar fields. Solid line histogram with 1$\sigma$ error denotes the
  number counts averaged in all the six fields. We also overlaid the
  number count of field galaxies evaluated in Subaru Deep Field
  (Maihara et al. 2001). We adopted {\it best} magnitude (see Source
  Extractor program manual; Bertin \& Arnouts 1996) in this plot, to
  compare them directly with the Maihara et al.'s result.}
\end{figure*}

Thirdly, it is also possible that our targetting fields did not cover
star forming galaxies by chance. We calculated field coverage
fractions (i.e., fraction of the area covered by our observations to
the area of the pair quasar field\footnote[2]{This means the area of
  circles on which all pair/triple quasars are located. For example,
  this area would be evaluated to be $\pi$$\times$(150\pp)$^2$ for
  Q2343/Q2344 pair quasars that are separated by 300\pp\ from each
  other on the sky. Areas used here are the lower limit of the size of
  possible cluster (group) of galaxies, because member galaxies
  probably distribute outside of this circles if they really exist.}).
We got effectively large values as $\sim$ 75\%\ and 33\%\ for
KP76/KP77/KP78 and Q2343/Q2344 fields, respectively, while the
coverage fraction for Q0301/Q0302 field is very small, $\sim$2\%. 
It seems unlikely that we happened to miss both star forming galaxies
and a number excess of galaxies in the fields around KP76/KP77/KP78 and
Q2343/Q2344 pair/triple quasars, if cluster of galaxies really
exist. In the case of Q0301/Q0302 field, the result is still open to
some uncertainties (e.g., coverage fraction could be underestimated,
if galaxies are distributed not spherically but filamentally in the
scale of $\sim$5Mpc along the pair quasars).

Finally, perhaps, most of \civ\ lines, whose counterparts were not
identified, could arise in star forming pockets outside bright
galaxies (e.g., the ejecta of Type Ia supernovae, dwarf galaxies, or
low surface brightness galaxies), which have already been suggested as
counterparts of weak ($W_{rest}$ $<$ 0.3 \AA) \mgii\ lines (Rigby et
al. 2002). These pockets have multiple phases; kiloparsec scale higher
ionization phase to produce \civ, and compact scale ($\sim$ 10pc) low
ionization phase with high density to produce \mgii\ (Charlton et
al. 2003). The redshift path density of weak \mgii\ systems is twice
of the strong \mgii\ systems that have almost always corresponding
luminous galaxies (Steidel 1995). Weak \mgii\ systems also have high
metallicity; solar or super-solar (Charlton et al. 2003). Nonetheless,
corresponding luminous galaxies are rarely found within
$\sim$50$h^{-1}$ kpc of the quasars (Churchill et al. 1999). All the
properties of weak \mgii\ systems are consistent with the
no-detection of star forming galaxies in our observations.

With $K$-band imaging observations, we placed photometric and $SFR$
upper limits of \ha\ emitters corresponding to \civ\ absorption lines
toward the lines of sight of pair/triple quasars. Deeper imaging
surveys for wider fields that cover completely pair/triple quasar
fields are necessary before concluding whether galaxy clusterings
corresponding to metal absorption lines really exist or not. A large
number of multiple quasars also have recently been discovered by the
large scale surveys such as Sloan Digital Sky Survey (SDSS; Schneider
et al. 2005) and 2dF QSO Redshift Survey (2QZ; Croom et al. 2004),
which enable us to extend this survey in the future.
\\

\acknowledgments
Part of the observation was kindly carried out in the engineering time
of Subaru Telescope. We would like to thank J.C. Charlton and
M. Eracleous for their comments to develop this study, and I. Tanaka
and M. Kajisawa for their useful advices on reduction of infrared
data. We also would like to thank D. Tytler and V. D'Odorico for
informing us of the coordinate of one of our targets. Finally, we wish
to thank the anonymous referee for many helpful comments and
suggestions.

\clearpage

% Table 1
\begin{deluxetable}{ccccc}
\tabletypesize{\scriptsize}
\tablecaption{Transmission of narrow band filters \label{tbl-1}}
\tablewidth{0pt}
\tablehead{
\colhead{(1)} & 
\colhead{(2)} & 
\colhead{(3)} &
\colhead{(4)} & 
\colhead{(5)} \\
\colhead{Filter} & 
\colhead{$\lambda_{cen}$ \tablenotemark{a}} & 
\colhead{$\Delta\lambda$ \tablenotemark{b}} & 
\colhead{$z(H\alpha)$ \tablenotemark{c}} &
\colhead{$\Delta \lambda_{NB}$/$\Delta \lambda_{K^{\prime}}$ \tablenotemark{d}} \\
\colhead{} & 
\colhead{($\mu m$)} & 
\colhead{($\mu m$)} &
\colhead{} & 
\colhead{(\%)} \\
}
\startdata
2.122       \tablenotemark{e} & 2.119 & 0.031 & 2.229 &  9.1 \\
2.248S      \tablenotemark{e} & 2.253 & 0.035 & 2.433 & 10.3 \\
$H_{2}(2-1)$\tablenotemark{f} & 2.250 & 0.022 & 2.428 &  6.9 \\
$H_{2}(1-0)$\tablenotemark{f} & 2.120 & 0.020 & 2.230 &  6.3 \\
\enddata
\tablenotetext{a}{Central wavelength of the filter.}
\tablenotetext{b}{Bandpass width of the filter with 50 \%
  transmission.}
\tablenotetext{c}{Redshift of a star-forming galaxy whose H$\alpha$
  emission line is redshifted into the center of the filter.}
\tablenotetext{d}{Bandpass ratio of narrow-band to \kp\ band filters
  in percent.}
\tablenotetext{e}{Filter available with UKIRT$+$UFTI.}
\tablenotetext{f}{Filter available with Subaru$+$CISCO.}
\end{deluxetable}

% Table 3
\addtocounter{table}{+1}
\begin{deluxetable}{ccccccc}
\tabletypesize{\scriptsize}
\tablecaption{Candidates for \ha\ emitters \label{tbl-3}}
\tablewidth{0pt}
\tablehead{
\colhead{(1)} & 
\colhead{(2)} & 
\colhead{(3)} &
\colhead{(4)} &
\colhead{(5)} & 
\colhead{(6)} \\
\colhead{Field} & 
\colhead{Object} & 
\colhead{\kp} & 
\colhead{$f_{H\alpha}$ \tablenotemark{a}} & 
\colhead{SFR \tablenotemark{b}} &
\colhead{FWHM} \\
\colhead{} & 
\colhead{} & 
\colhead{(mag)} &
\colhead{(ergs s$^{-1}$ cm$^{-2}$)} & 
\colhead{($h^{-2}$M$_{\odot}$ yr$^{-1}$)} &
\colhead{(\pp)} \\
}
\startdata
Q2343$+$125 & A & 14.92 & 1.35$\times$$10^{-15}$ & 466 & 0.97 \\
            & B & 16.87 & 2.17$\times$$10^{-16}$ &  75 & 0.99 \\
Q2344$+$125 & C & 16.29 & 3.14$\times$$10^{-16}$ & 108 & 0.82 \\
            & D & 17.64 & 9.34$\times$$10^{-17}$ &  32 & 0.80 \\
            & E & 18.28 & 5.67$\times$$10^{-17}$ &  20 & 0.65 \\
\enddata
\tablenotetext{a}{Total flux of \ha\ emission line, without a
  correction for the effects of [\nii] line.}
\tablenotetext{b}{Star formation rate in unit of solar-mass per year.}
\end{deluxetable}

% Table 2
\addtocounter{table}{-2}
\begin{deluxetable}{llllccrcccccc}
\rotate
\tabletypesize{\scriptsize}
\setlength{\tabcolsep}{0.05in}
\tablecaption{Observation Log \label{tbl-2}}
\tablewidth{0pt}
\tablehead{
\colhead{(1)} & 
\colhead{(2)} & 
\colhead{(3)} &
\colhead{(4)} & 
\colhead{(5)} & 
\colhead{(6)} & 
\colhead{(7)} &
\colhead{(8)} & 
\colhead{(9)} &
\colhead{(10)} &
\colhead{(11)} &
\colhead{(12)} &
\colhead{(13)} \\
\colhead{QSO} &
\colhead{\zem} &
\colhead{\zabs} &
\colhead{Ion} &
\colhead{Filter} & 
\colhead{Date} &
\colhead{Exposure} &
\colhead{Seeing} &
\colhead{3$\sigma$} &
\colhead{5$\sigma$} &
\colhead{\fha($lim.$)$^{a}$} &
\colhead{SFR($lim.$)$^{b}$} &
\colhead{Reference} \\
\colhead{} & 
\colhead{} & 
\colhead{} &
\colhead{} &
\colhead{} & 
\colhead{} & 
\colhead{(sec)} & 
\colhead{(\pp)} &
\colhead{(mag)} & 
\colhead{(mag)} &
\colhead{(\flux)} & 
\colhead{($h^{-2}$\spy)} &
\colhead{} \\
}
\startdata
KP76            $^{c}$ & 2.467       & 2.2462        & \civ, \siiv & K98          & May 24--25 2002 & 11760 & 0.42 & 21.39 & 21.03 & $2.4\times10^{-17}$ & 6.8 & 1 \\
                       &             &               &             & 2.122        & May 24--25 2002 & 33100 & 0.42 & 20.91 & 20.39 &                     &     &   \\
KP77/KP78       $^{d}$ & 2.526,2.605 & 2.2445,2.2417 & \civ, \siiv & \kp\         & May   14   2003 &  1440 & 0.40 & 21.09 & 20.42 & $6.3\times10^{-18}$ & 1.8 & 1 \\
                       &             &               &             & $H_{2}(1-0)$ & May   14   2003 &  5760 & 0.39 & 20.00 & 19.34 &                     &     &   \\
Q0301$-$005     $^{c}$ & 3.223       & 2.4291        & \civ, \siiv & K98          & Dec 22--23 2002 &  4860 & 0.82 & 20.06 & 19.34 & $6.9\times10^{-16}$ & 236 & 2 \\
                       &             &               &             & 2.248S       & Dec 22--23 2002 & 11700 & 0.88 & 18.77 & 18.14 &                     &     &   \\
Q0302$-$003     $^{c}$ & 3.285       & 2.4233        & \civ, \siiv & K98          & Dec   22   2002 &  4860 & 0.48 & 21.07 & 20.44 & $6.7\times10^{-17}$ &  23 & 2 \\
                       &             &               &             & 2.248S       & Dec   22   2002 & 12600 & 0.50 & 20.15 & 19.59 &                     &     &   \\
Q2343$+$125     $^{d}$ & 2.515       & 2.4285,2.4308 & \civ        & \kp\         & Nov   19   2000 &  3260 & 0.94 & 20.54 & 20.03 & $1.6\times10^{-17}$ & 5.5 & 3 \\
                       &             &               &             & $H_{2}(2-1)$ & Nov   19   2000 &  8640 & 0.93 & 19.21 & 18.51 &                     &     &   \\
Q2344$+$125     $^{d}$ & 2.763       & 2.4265,2.4292 & \civ        & \kp\         & Nov    7   2003 &  1920 & 0.52 & 21.14 & 20.52 & $9.3\times10^{-18}$ & 3.2 & 3 \\
                       &             &               &             & $H_{2}(2-1)$ & Nov    7   2003 &  5760 & 0.52 & 20.74 & 20.07 &                     &     &   \\
\enddata
\tablenotetext{a}{3$\sigma$ detection limit of \ha\ emission line.}
\tablenotetext{b}{3$\sigma$ detection limit of star formation rate.}
\tablenotetext{c}{Observed with UKIRT + UFTI.}
\tablenotetext{d}{Observed with Subaru + CISCO.}
\tablerefs{(1) Crotts \& Fang 1998;
           (2) Steidel 1990, Cowie et al. 1995; Dobrzycki \& Bechtold 1991;
           (3) Teplitz et al. 1998}
\end{deluxetable}


\begin{thebibliography}{}
\bibitem[Bergvall et al.(1997)]{ber97} Bergvall, N., \"Ostlin, G.,
  Karlsson, K. G., \"Orndahl, E., and R\"onnback, J., 1997, \aap,
  321, 771
\bibitem[Bergeron and Boiss\'e(1991)]{ber91} Bergeron, J., and
  Boiss\'e, P., 1991, \aap, 243, 344
\bibitem[Bertin and Arnouts (1996)]{ber96} Bertin, E., and Arnouts,
  S., 1996, \aaps, 117, 393
\bibitem[Bunker et al.(1999)]{bun99} Bunker, A.J., Warren, S.J.,
  Clements, D.L., Williger, G.M., and Hewett, P.C., 1999, \mnras, 309,
  875
\bibitem[Charlton et al.(2003)]{cha03} Charlton, J.C., Ding, J.,
  Zonak, S.G., Churchill, C.W., Bond, N.A., Rigby, J.R., 2003, \apj,
  589, 111
\bibitem[Charlton and Churchill(1996)]{char96} Charlton, J. C. and
  Churchill, C. W., 1996, \apj, 465, 631
\bibitem[Churchill et al.(1999)]{chur99} Churchill, C. W., Rigby, J. R.,
  Charlton, J. C., and Vogt, S. S, 1999, \apjs, 120, 51
\bibitem[Chen et al.(2001)]{cen01a} Chen, H.-W., Lantezza, K. M., and
  Webb, 2001, \apj, 556, 158
\bibitem[Chen et al.(2001)]{cen01b} Chen, H.-W., Lantezza, K. M.,
  Webb, J. K., and Barcons, X., 2001, \apj, 559, 654
\bibitem[Cowie et al.(1995)]{cow95} Cowie, L.L., Songaila, A., Kim,
  T.-S., and Hu, E.M., 1995, \aj, 109, 1522
\bibitem[Croom et al.(2004)]{cro04} Croom, S. M., Smith, R. J., Boyle,
  B. J., Shanks, T., Miller, L., Outram, P. J., and Loaring, N. S.,
  2004, \mnras, 349, 1397
\bibitem[Crotts and Fang(1998)]{cro98} Crotts, A. P. S., and Fang, Y.,
  1998, \apj, 502, 16 
\bibitem[Dobrzycki et al.(1991)]{dob91} Dobrzycki, A., and Bechtold,
  J., 1991, \apj, 377, 69
\bibitem[Francis and Hewett(1993)]{fra93} Francis, P. J., and Hewett,
  P. C., 1993, \aj, 105, 1633
\bibitem[Francis et al.(1996)]{fra96} Francis, P. J., Woodgate, B. E.,
  Warren, S. J., M\o ller, P., Mazzolini, M., Bunker, A. J.,
  Lowenthal, J. D., Williams, T. B., Minezaki, T., Kobayashi, Y., and
  Yoshii, Y., 1996, \apj, 457, 490
\bibitem[Francis, Woodgate, and Danks(1997)]{fra97} Francis, P. J.,
  Woodgate, B., and Danks, A. C., 1997, \apjl, 482, 25
\bibitem[Goodrich, Veilleux, and Hill(1994)]{goo94} Goodrich, R.W.,
  Veilleux, S., and Hill, G.J., 1994, \apj, 422, 521
\bibitem[Hu and McMahon(1996)]{hu96} Hu, E.M., and McMahon, R.G.,
  1996, \nat, 382, 231
\bibitem[Iwamuro et al.(2000)]{iwa00} Iwamuro, F., Motohara, K.,
  Maihara, T., Iwai, J., Tanabe, H., Taguchi, T., Hata, R., Terada,
  H., Goto, M., Ohya, S., Iye, M., Yoshida, M., Karoji, H., Ogasawara,
  R., and Sekiguchi, K., 2000, \pasj, 52, 73
\bibitem[Iye et al.(2004)]{iye04} Iye, M., et al., 2004, \pasj, 56,
  381
\bibitem[Jakobsen et al.(1986)]{jak86} Jakobsen, P., Perryman,
  M. A. C., Ulrich, M. H., Macchetto, F., and di Serego Alighieri,
  S., 1986, \apjl, 303, 27
\bibitem[Juneau et al.(2005)]{jun05} Juneau, S., et al., 2005, \apjl,
  619, 135
\bibitem[Kawara, Nishidat, Taniguchi(1988)]{kaw88} Kawara, K.,
  Nishida, M., and Taniguchi, Y., 1988, \apj, 328, L41
\bibitem[Kennicutt (1998)]{ken98} Kennicutt, R. C. Jr., 1998, \araa,
  36, 189
\bibitem[Lanzetta et al.(1998)]{lan98} Lanzetta, K. M., Chen, H.-W.,
  Webb, J. K., and Barcons, X., 1998, invited review for IAU
  Colloquium 171, The Low Surface Brightness Universe
\bibitem[LeFevre et al. (1996)]{lef96} LeF$\grave{e}$vre, O.L.,
  Deltorn, J.M., Crampton, D., and Dickinson, M., 1996, \apjl, 471,
  11
\bibitem[Lu, Sargent, and Barlow(1998)]{lu98} Lu, L., Sargent, W.L.W.,
  and Barlow, T.A., 1998, \aj, 115, 55
\bibitem[Maihara et al.(2001)]{mai01} Maihara, T. et al., 2001, PASJ,
  53, 25
\bibitem[Moorwood et al.(2000)]{moo00} Moorwood, A.F.M., van der Werf,
  P.P., Cuby, J.G., and Oliva, E., 2000, \aap, 362, 9
\bibitem[Motohara et al.(1998)]{mot98} Motohara, K., et al., 1998, in
  Proc. SPIE 3354: Infrared Astronomical Instrumentation,
  ed. A. M. Fowler, 659
\bibitem[Pascarelle et al. (1996)]{pas96} Pascarelle, S.M., Windhorst,
  R.A., Driver, S.P., Ostrander, E.J., and Keel, W.C., 1996, \apjl,
  456, 21
\bibitem[Pentricci et al.(2000)]{pen00} Pentricci, L., Kurk, J. D.,
  R\"ottgering, H. J. A., Miley, G. K., van Breugel, W., Carilli,
  C. L., Ford, H., Heckman, T., McCarthy, P., and Moorwood, M., 2000,
  \aap, 361, L25
\bibitem[Pettini et al.(1998)]{pet98} Pettini, M., Kellogg, M.,
  Steidel, C.C., Dickinson, M., Adelberger, K.L., and Giavalisco, M.,
  1998, \apj, 508, 539
\bibitem[Rigby, Charlton, and Churchill(2002)]{rig02} Rigby, J.R.,
  Charlton, J.C., and Churchill, C., 2002, \apj, 565, 743
\bibitem[Roche et al.(2002)]{roc02} Roche P.F. et al. 2002, Proc Spie
  4841, Instrument Design and Performance for Optical/IR Ground-Based
  Telescopes, eds. M Iye and A.F Moorwood
\bibitem[Sargent, Boksenberg, and Steidel(1988)]{sar88} Sargent,
  W.L.W., Boksenberg, A., and Steidel, S.S., 1988, \apjs, 68, 539
\bibitem[Schneider et al.(2005)]{sch05} Schneider, D. P., et al.,
  2005, astro-ph/0505679
\bibitem[Shaver, Boksenberg, and Robertson(1982)]{sha82} Shaver,
  P. A., Boksenberg, A., and Robertson, J. G., 1982, \apjl, 261, 7
\bibitem[Simpson et al.(1996)]{sim96} Simpson, C., Forbes, D.A.,
  Baker, A.C., and Ward, M.J., 1996, \mnras, 283, 777
\bibitem[Smette et al.(1995)]{sme95} Smette, A., Robertson, J.G.,
  Shaver, P.A., Reimers, D., Wisotzki, L., and Koehler, T., 1995,
  \aaps, 113, 199
\bibitem[Songaila(1998)]{son98} Songaila, A., 1998, \aj, 115, 2184
\bibitem[Steidel(1995)]{ste95} Steidel, C.C., 1995, in QSO Absorption
  Lines, ed. G. Meylan (Garching: Springer Verlag), 139
\bibitem[Steidel, Dickinson, and Persson(1994)]{ste94} Steidel, C. C.,
  Dickinson, M., and Persson, S. E., \apjl, 437, 75
\bibitem[Steidel(1990)]{ste90} Steidel, C.C., 1990, \apjs, 74, 37
\bibitem[Teplitz, Malkan, and McLean(1998)]{tep98} Teplitz. H.I.,
  Malkan, M., and McLean, I. S., 1998, \apj, 506, 519
\bibitem[Tamura et al.(2001)]{tam98} Tamura, N., Ohta, K., Maihara,
  T., Iwamuro, F., Motohara, K., Takata, T., and Iye, M., 2001,
  \pasj, 53, 653
\bibitem[Warren et al.(1994)]{war94} Warren, S. J., Hewett, P. C., and
  Osmer, P. S., 1994, \apj, 421, 412
\end{thebibliography}
\end{document}